\documentclass[sigconf]{acmart}

\usepackage{balance}
\usepackage{multirow}
\usepackage{subcaption}
\usepackage{enumitem}

\AtBeginDocument{%
  }

\copyrightyear{2026}
\acmYear{2026}
\setcopyright{cc}
\setcctype{by-nc-nd}
\acmConference[SIGIR '26]{Proceedings of the 49th International ACM SIGIR Conference on Research and Development in Information Retrieval}{July 20--24, 2026}{Melbourne, VIC, Australia}
\acmBooktitle{Proceedings of the 49th International ACM SIGIR Conference on Research and Development in Information Retrieval (SIGIR '26), July 20--24, 2026, Melbourne, VIC, Australia}
\acmDOI{10.1145/3805712.3808581}
\acmISBN{979-8-4007-2599-9/2026/07}

\begin{document}

%%
%% The "title" command has an optional parameter,
%% allowing the author to define a "short title" to be used in page headers.
\title{A Sketch+Text Composed Image Retrieval Dataset for Thangka}

%%
%% The "author" command and its associated commands are used to define
%% the authors and their affiliations.
%% Of note is the shared affiliation of the first two authors, and the
%% "authornote" and "authornotemark" commands
%% used to denote shared contribution to the research.
\author{Jinyu Xu}
\orcid{0000-0002-9246-4800}
\affiliation{%
  \institution{Wuhan University of Technology}
  % \department{School of Computer Science and Artificial Intelligence}
  \city{Wuhan}
  \state{Hubei}
  \country{China}
}
\email{jinyxu@whut.edu.cn}

\author{Yi Sun}
\orcid{0009-0009-7383-7809}
\affiliation{%
  \institution{Wuhan University of Technology}
  \city{Wuhan}
  \state{Hubei}
  \country{China}
}
\email{syi1005@whut.edu.cn}

\author{Jiangling Zhang}
\orcid{0009-0007-1219-7807}
\affiliation{%
  \institution{Wuhan University of Technology}
  \city{Wuhan}
  \state{Hubei}
  \country{China}
}
\email{zhangjiangling@whut.edu.cn}

\author{Qing Xie}
\orcid{0000-0003-4530-588X}
\authornote{Corresponding authors.}
\affiliation{%
  \institution{Wuhan University of Technology}
  \city{Wuhan}
  \state{Hubei}
  \country{China}
}
\email{felixxq@whut.edu.cn}

\author{Daomin Ji}
\orcid{0009-0000-0037-3614}
\affiliation{%
  \institution{RMIT University}
  \city{Melbourne}
  \country{Australia}
}
\email{daomin.ji@student.rmit.edu.au}

\author{Zhifeng Bao}
\orcid{0000-0003-2477-381X}
\affiliation{%
  \institution{The University of Queensland}
  \city{Brisbane}
  \country{Australia}
}
\email{zhifeng.bao@uq.edu.au}

\author{Jiachen Li}
\orcid{0000-0002-0602-9360}
\authornotemark[1]
\affiliation{%
  \institution{Wuhan University of Technology}
  \city{Wuhan}
  \state{Hubei}
  \country{China}
}
\email{lijiachen@whut.edu.cn}

\author{Yanchun Ma}
\orcid{0009-0001-4600-749X}
\affiliation{%
  \institution{Wuhan Vocational College of Software and Engineering}
  \city{Wuhan}
  \state{Hubei}
  \country{China}
}
\email{mayanchun@whvcse.edu.cn}

\author{Yongjian Liu}
\orcid{0009-0005-8388-6736}
\affiliation{%
  \institution{Wuhan University of Technology}
  \city{Wuhan}
  \state{Hubei}
  \country{China}
}
\email{liuyj@whut.edu.cn}

%%
%% By default, the full list of authors will be used in the page
%% headers. Often, this list is too long, and will overlap
%% other information printed in the page headers. This command allows
%% the author to define a more concise list
%% of authors' names for this purpose.
\renewcommand{\shortauthors}{Jinyu Xu et al.}

%%
%% The abstract is a short summary of the work to be presented in the
%% article.
\begin{abstract}
Composed Image Retrieval (CIR) enables image retrieval by combining multiple query modalities, but existing benchmarks predominantly focus on general-domain imagery and rely on reference images with short textual modifications.
As a result, they provide limited support for retrieval scenarios that require fine-grained semantic reasoning, structured visual understanding, and domain-specific knowledge.
In this work, we introduce \textbf{CIRThan}, a sketch+text \textbf{C}omposed \textbf{I}mage \textbf{R}etrieval dataset for \textbf{Than}gka imagery, a culturally grounded and knowledge-specific visual domain characterized by complex structures, dense symbolic elements, and domain-dependent semantic conventions.
CIRThan contains 2,287 high-quality Thangka images, each paired with a human-drawn sketch and hierarchical textual descriptions at three semantic levels, enabling composed queries that jointly express structural intent and multi-level semantic specification.
We provide standardized data splits, comprehensive dataset analysis, and benchmark evaluations of representative supervised and zero-shot CIR methods.
Experimental results reveal that existing CIR approaches, largely developed for general-domain imagery, struggle to effectively align sketch-based abstractions and hierarchical textual semantics with fine-grained Thangka images, particularly without in-domain supervision.
We believe CIRThan offers a valuable benchmark for advancing sketch+text CIR, hierarchical semantic modeling, and multimodal retrieval in cultural heritage and other knowledge-specific visual domains.
The dataset is publicly available at \url{https://github.com/jinyuxu-whut/CIRThan}.
\end{abstract}

%%
%% The code below is generated by the tool at http://dl.acm.org/ccs.cfm.
%% Please copy and paste the code instead of the example below.
%%


%%
%% Keywords. The author(s) should pick words that accurately describe
%% the work being presented. Separate the keywords with commas.

% \received{20 February 2007}
% \received[revised]{12 March 2009}
% \received[accepted]{5 June 2009}

%%
%% This command processes the author and affiliation and title
%% information and builds the first part of the formatted document.
\maketitle

\section{Introduction}\label{sec:intro}

With the increasing demand for fine-grained image retrieval, users rely on multimodal queries to express search intents that cannot be captured by a single modality alone.
To address this need, composed image retrieval (CIR)~\cite{vo2019composing,liu2021image,baldrati2022conditioned,saito2023pic2word,baldrati2023circo,sun2023training,sun2026sdrcir} has emerged as an important paradigm, typically combining a reference image with a relative textual description.
While most existing CIR studies focus on general-domain natural images, users often encounter retrieval scenarios that require domain-specific knowledge and detailed semantic reasoning, posing challenges to existing CIR benchmarks.
A representative example is Thangka imagery~\cite{wang2021thanka, li2026guided}, a form of traditional Tibetan art characterized by highly structured compositions, dense elements, and extremely fine textures along with profound cultural meanings.
As an important component of cultural heritage and digital humanities, Thangka imagery plays a crucial role in the digital organization, analysis, and dissemination of cultural collections~\cite{hu2025tovect, wang2021thanka, dong2023multi, ren2024dunhuang}.
This highlights the need for a CIR benchmark specifically designed for knowledge-specific visual domains such as Thangka.

Nevertheless, existing CIR benchmarks remain limited in several important aspects.

\begin{figure}[thbp]
    \centering
    \includegraphics[width=\linewidth]{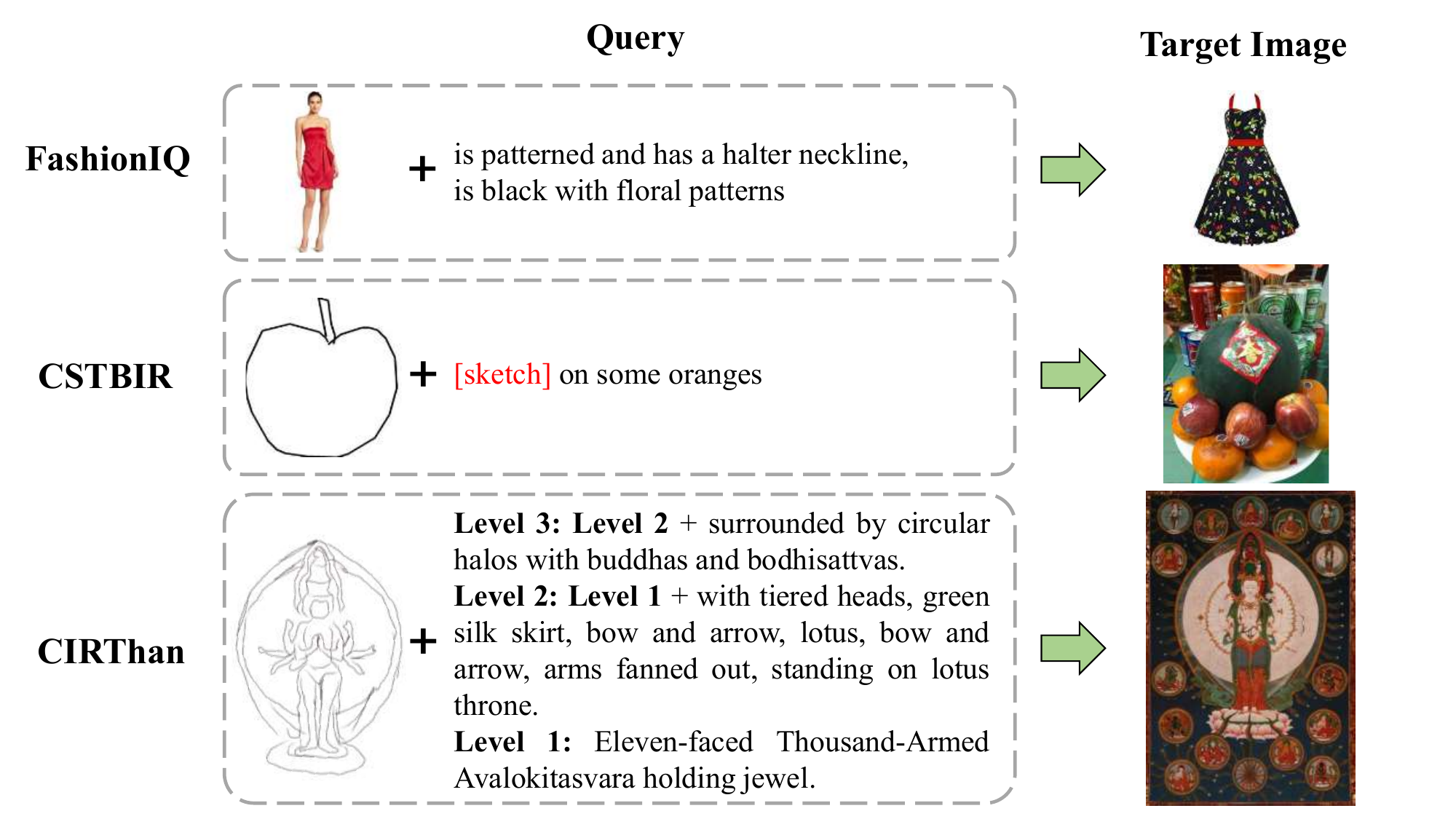}
    \caption{Comparison of composed query examples from FashionIQ, CSTBIR, and the proposed CIRThan dataset.}
    \label{fig:1}
\end{figure}

\textbf{(1) Limitation of the visual reference.}
Existing CIR benchmarks such as FashionIQ~\cite{wu2021fashionIQ}, CIRR~\cite{liu2021image}, and CIRCO~\cite{baldrati2023circo} are constructed on general-domain image collections, where images depict common objects with loosely structured and readily interpretable visual semantics (Figure~\ref{fig:1}).
Accordingly, conventional CIR assumes that users can identify a suitable reference image that is already semantically close to the desired target and express the remaining differences using a short textual description.
However, these assumptions become fragile when applied to complex scenarios or domain-specific visual collections such as Thangka imagery, where identifying an appropriate reference image is inherently difficult.
As an alternative, sketches provide an intuitive and flexible way for users to express visual intent, allowing them to abstract away irrelevant details and focus on salient structures and shape cues.
While sketches have been explored as an alternative reference modality, existing sketch+text CIR benchmarks (e.g., CSTBIR~\cite{gatti2024composite} in Figure~\ref{fig:1}) typically employ coarse or category-level sketches that mainly convey object presence.
Such sketches are insufficient to support fine-grained composed image retrieval, where detailed structural configurations and subtle visual semantic cues are critical.

% Such sketches are insufficient for fine-grained retrieval intent, where discriminative cues are embedded in fine-grained structural configurations and high-density semantic details, limiting the effectiveness of existing sketch-based CIR formulations.

\textbf{(2) Limitation of the textual description.}
Textual descriptions in existing CIR benchmarks~\cite{wu2021fashionIQ, liu2021image, baldrati2023circo} are primarily designed for general-domain imagery and typically focus on coarse semantic cues that can be easily expressed in brief natural language.
As a result, they provide limited support for retrieval scenarios that require domain-specific and fine-grained semantic descriptions, such as those involving complex structures, instance-level attributes, or knowledge-dependent semantic details.
Furthermore, in practical retrieval scenarios, the level of semantic detail expressed in textual queries often varies across users, reflecting differences in domain knowledge and retrieval intent.
However, current CIR datasets do not explicitly model the hierarchical nature of textual descriptions, which further restricts their applicability to knowledge-specific and fine-grained composed image retrieval.

These limitations reveal that existing CIR benchmarks are insufficient to support composed image retrieval scenarios where satisfying retrieval intent requires precise semantic matching between the composed query and target images, particularly in knowledge-specific scenarios that demand domain-aware semantics, detailed structural understanding, and multi-level semantic specification.
Although sketch+text composed image retrieval has recently attracted initial attention, existing studies remain limited, leaving sketch+text CIR insufficiently explored as a benchmark setting for such demanding retrieval scenarios.

Motivated by these limitations, we introduce \textbf{CIRThan}, a new sketch+text \textbf{C}omposed \textbf{I}mage \textbf{R}etrieval dataset for \textbf{Than}gka imagery.
To address \textbf{Limitation (1)}, CIRThan adopts fine-grained human-drawn sketches as the visual reference modality, enabling users to convey salient objects, structural layouts, and shape-related cues that are difficult to express using a semantically close natural image or a coarse category-level sketch.
To address \textbf{Limitation (2)}, CIRThan provides hierarchical textual descriptions grounded in domain knowledge and retrieval intent, explicitly modeling different levels of semantic granularity to accommodate user variability in domain knowledge and expressive detail.
Overall, fine-grained sketches and hierarchical text form an expressive query formulation tailored to CIR in knowledge-specific visual domains.

In summary, our contributions are:
(\textbf{i}) We introduce \textbf{CIRThan}, a new sketch+text composed image retrieval dataset for Thangka imagery, which serves as a knowledge-specific benchmark for evaluating composed retrieval under strict semantic matching requirements;
(\textbf{ii}) We design a curated data construction pipeline that integrates fine-grained human-drawn sketches with hierarchical textual descriptions grounded in domain knowledge, enabling composed queries that capture detailed visual intent and multi-level semantic specification;
(\textbf{iii}) We conduct extensive benchmarks of representative supervised and zero-shot CIR methods on CIRThan, revealing several quantitative findings: 
1) a substantial performance gap between supervised and zero-shot CIR, with the best supervised method achieving 52.03\% R@1 while zero-shot methods remain below 8\% R@1 on average; 
2) consistent performance gains with increased textual granularity across all methods, highlighting the importance of hierarchical semantic specification; and 
3) limited effectiveness of current MLLM-based zero-shot CIR approaches in aligning sketch+text composed queries with domain-specific semantics, underscoring the challenges of fine-grained CIR in knowledge-specific visual domains.

\section{Related Work}\label{sec:related_work}

\subsection{Sketch-based \& Text-based Image Retrieval}
Sketch-based image retrieval (SBIR) and text-based image retrieval (TBIR) are two classical paradigms that allow users to express retrieval intent through a single modality.
In SBIR~\cite{Qi2016Siamese, Seddati2017Triplet, BUI2018hybrid, collomosse2019livesketch, Lei2020Semi3Net, jinyu2026enhancing}, sketches provide a flexible and human-centered visual abstraction that emphasizes shape, contour, and structural cues, making them effective when retrieval intent is primarily defined by visual structure rather than detailed appearance~\cite{sun2022dlinet, ling2022mlrm, sain2023sketchpvt, zheng2021DeepParsing}.
However, sketches alone provide limited access to explicit semantic information, such as object identity, attributes, or symbolic meaning, which restricts their applicability in finer-grained matching scenarios.

TBIR~\cite{lulf2024clip-braches, huo2023deep, jiang2023cross} enables semantic matching between images and natural language descriptions by learning joint visual–textual representations.
Although recent vision–language models~\cite{radford2021learning, jia2021scaling, yao2022filip} have substantially improved TBIR performance in general-domain settings, text alone often struggles to convey structural layout, spatial relations, or domain-specific visual conventions.
As a result, neither sketches nor text alone is sufficient to express complex retrieval intent that requires both structural precision and rich semantic understanding, motivating composed image retrieval.

\subsection{Composed Image Retrieval}
Composed Image Retrieval (CIR) extends single-modality retrieval by combining a reference image with a textual description to retrieve a target image that satisfies a specified modification.
Existing CIR methods can be broadly categorized into supervised methods~\cite{baldrati2022conditioned, baldrati2022effective} trained on annotated triplets and zero-shot methods~\cite{saito2023pic2word, suo2024keds, wang2025Generative} that leverage weakly labeled image-text data or pretrained vision-language models, with recent work further exploring training-free settings using large language models~(LLMs)~\cite{yang2024semantic, sun2025cotmr, karthik2024vision}.

Most existing CIR benchmarks, including FashionIQ~\cite{wu2021fashionIQ}, CIRR~\cite{liu2021image}, CIRCO~\cite{baldrati2023circo}, and GeneCIS~\cite{vaze2023genecis}, are constructed on general-domain imagery and primarily rely on textual descriptions to specify attribute- or object-level modifications. In these datasets, reference images often provide limited fine-grained semantic guidance, and retrieval performance largely depends on textual cues rather than precise semantic correspondence between reference and target images. For instance, CIRR~\cite{liu2021image} constructs reference-target pairs based on visual similarity without careful human verification and subsequently annotates semantic differences using text, which may result in reference–target image pairs that differ substantially from each other. As a consequence, retrieval often relies more heavily on textual cues than on precise semantic correspondence between the reference and the target, limiting the evaluation of fine-grained CIR.

To alleviate the difficulty of acquiring suitable reference images, sketch+text CIR has recently been explored, allowing users to express visual intent more flexibly through sketches.
However, existing sketch+text benchmarks~\cite{koley2024youll, gatti2024composite} predominantly employ coarse or category-level sketches and focus on general-domain scenarios, leaving fine-grained structural semantics and knowledge-specific domains largely underexplored.
In contrast, our work investigates sketch+text CIR in a knowledge-specific visual domain, emphasizing fine-grained sketches and hierarchical textual specification to support more precise semantic alignment.

\section{The CIRThan Dataset}\label{sec:dataset}

This section introduces the proposed CIRThan dataset.
We first formalize the sketch+text composed image retrieval~(CIR) task and then detail the data construction pipeline.
Then, we present a comprehensive statistical analysis to characterize the dataset.
% Finally, we discuss potential research directions enabled by CIRThan.

\subsection{Sketch+Text CIR Task Definition}\label{sec:task_definition}

Let us define a composed query as $\{\mathcal{I}_s, \mathcal{T}\}$, where $\mathcal{I}_s$ denotes a reference sketch that conveys salient visual cues and fine-grained visual details that are difficult to articulate precisely using natural language, and $\mathcal{T}$ denotes a natural language description specifying semantic attributes and complementary information. 
Given an image database $\mathcal{D}$ consisting of Thangka images, the objective of the sketch+text composed image retrieval task is to retrieve a target image  $\mathcal{I}_t \in \mathcal{D}$ that precisely matches the visual intent conveyed by $\mathcal{I}_s$ while preserving the semantic content expressed in $\mathcal{T}$. 
This formulation is model-agnostic and serves as a benchmark setting for evaluating sketch+text composed image retrieval in knowledge-specific visual domains.

\subsection{Data Construction}

\begin{figure}
    \centering
    \includegraphics[width=\linewidth]{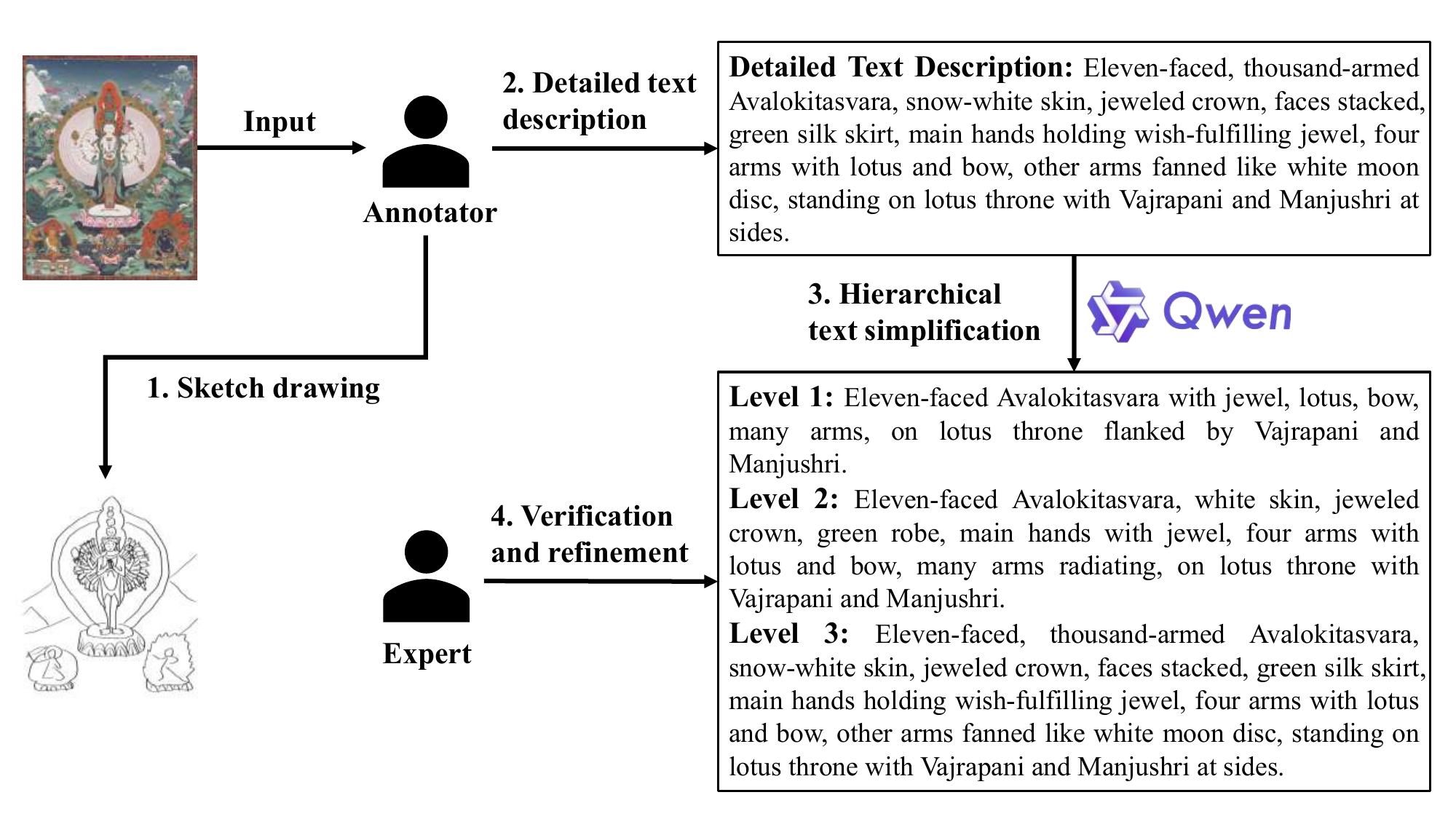}
    \caption{The data construction pipeline of the CIRThan dataset.}
    \label{dataflow}
\end{figure}

The Thangka images in CIRThan are provided by the \textit{Tibet Thangka Art and Culture Association}.
All images are obtained with permission from the source institutions or are publicly available under appropriate usage conditions.
The collection comprises high-quality photographs collected from monasteries, museums, and individual artists across Tibet, ensuring both cultural authenticity and visual diversity.

To construct a high-quality and semantically reliable CIR dataset, we design a standardized annotation pipeline that integrates human expertise with large language model (LLM) assistance.
As a culturally visual domain, Thangka imagery involves various characters, symbolic objects, and visual configurations rooted in Buddhist philosophy.
Accordingly, our annotation pipeline emphasizes terminological standardization, semantic precision, and cross-modal consistency across sketches, texts, and images.

The pipeline consists of four sequential steps:
(1) sketch drawing,
(2) detailed textual description,
(3) hierarchical text simplification, and
(4) verification and refinement.
Each Thangka image is paired with one human-drawn sketch and three hierarchical textual descriptions (Level~1--3).
Figure~\ref{dataflow} illustrates the overall workflow.

\textbf{Sketch Drawing.}
Annotators draw a free-form sketch to capture discriminative visual cues of the input Thangka, such as characteristic contours, distinctive shape patterns, local texture, and salient structural configurations.
Depending on the content, sketches may include coarse surrounding structures to convey global spatial organization.
Rather than exhaustively depicting all entities, sketches reflect a selective abstraction that aligns with realistic user behavior in composed retrieval.

\begin{figure*}[t]
    \centering
    \begin{subfigure}[t]{0.29\linewidth}
        \centering
        \includegraphics[width=\linewidth]{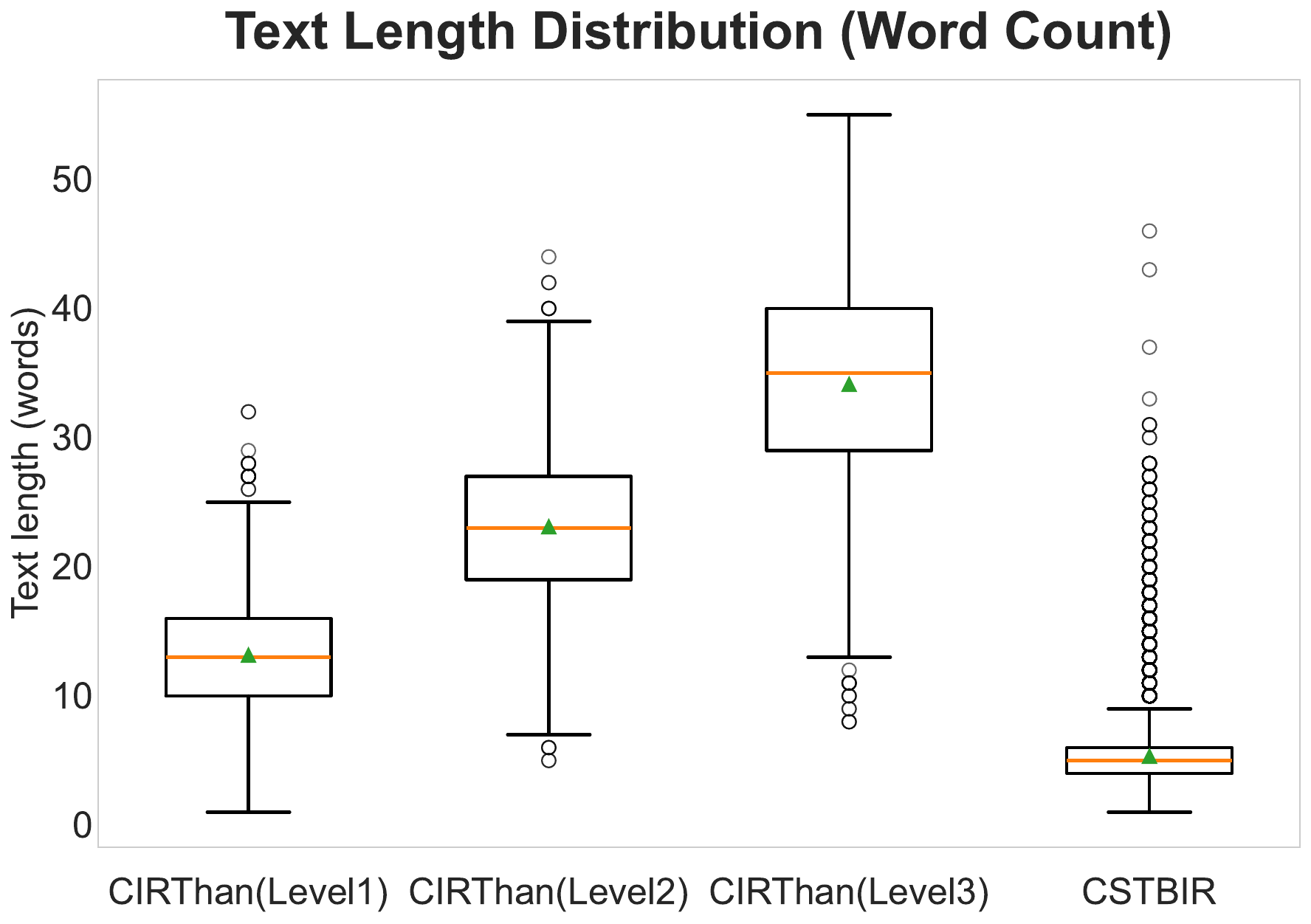}
        \caption{Distribution of text lengths}
        \label{text_length_box}
    \end{subfigure}
    \begin{subfigure}[t]{0.32\linewidth}
        \centering
        \includegraphics[width=\linewidth]{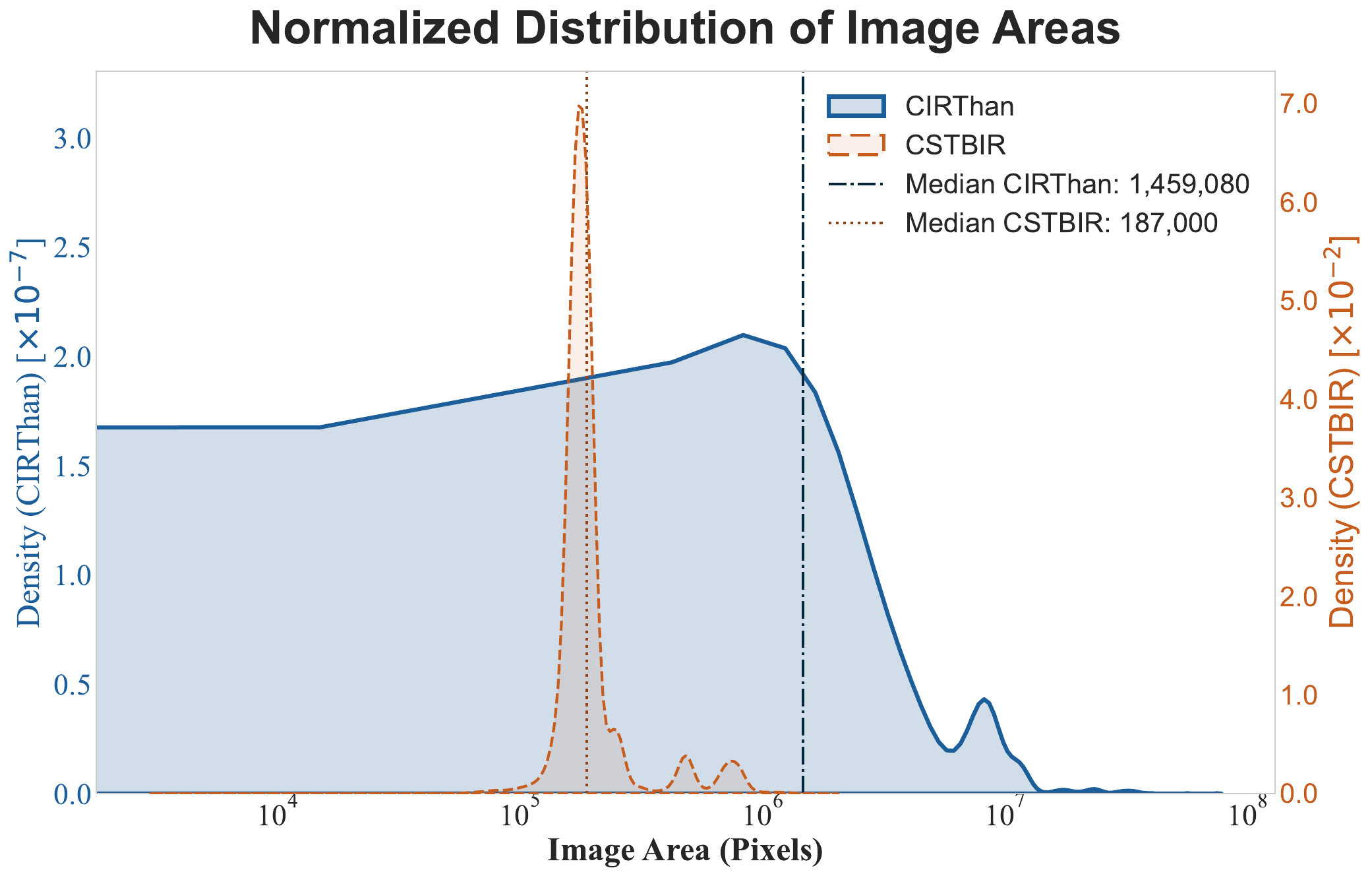}
        \caption{Image area distribution}
        \label{area_distribution}
    \end{subfigure}
    \begin{subfigure}[t]{0.32\linewidth}
        \centering
        \includegraphics[width=\linewidth]{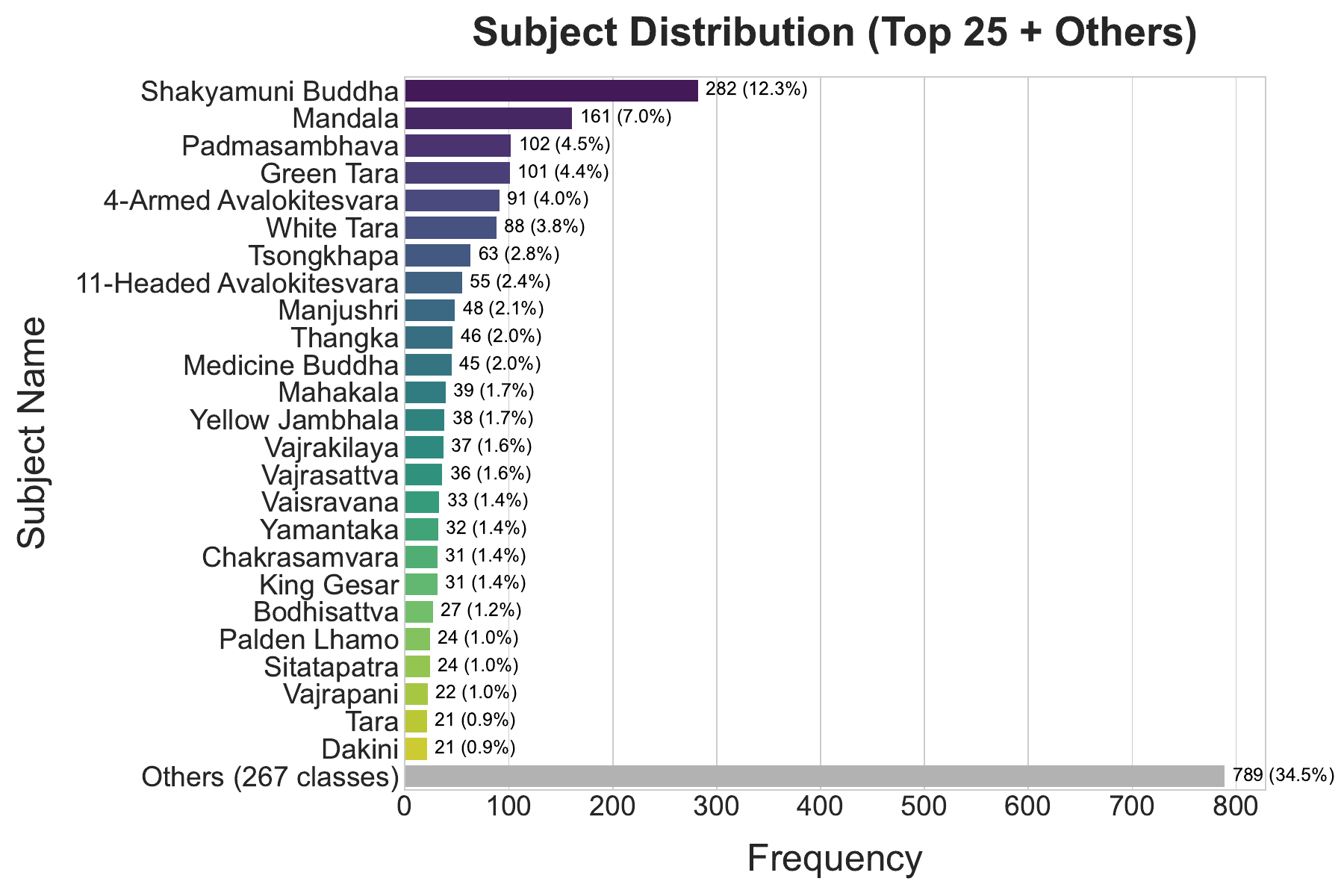}
        \caption{Subject distribution}
        \label{tangka_top30_horizontal}
    \end{subfigure}
    \caption{Statistical analysis of image resolution and subject distribution in the dataset.}
    \label{fig:statistic}
\end{figure*}

\textbf{Detailed Textual Description.}
To complement the sketch modality, annotators provide a detailed natural language description to specify the semantic content of salient entities depicted in Thangka imagery.
The description explicitly identifies key characters, symbolic objects, and other relevant visual elements, together with their attributes (e.g., color and ornamentation) and characteristics.
In addition, it captures contextual cues, semantic relationships and other semantic details that are difficult to convey reliably through sketches alone.
To ensure terminological accuracy and consistency, we provide annotation guidelines with standardized vocabulary and manage the labeling process using \textit{Label Studio}.

\textbf{Hierarchical Text Simplification.}
To model variability in user expertise and query specificity, we construct three levels of textual descriptions via controlled simplification.
This step does not introduce new semantic content; instead, it systematically reduces descriptive detail while preserving the core semantic structure of the original description.
We use Qwen-3 as an assisting LLM under strict constraints: the primary semantics must be retained, while simplification is limited to secondary attributes, decorative details, and redundant expressions.
Specifically, \textbf{Level~3} is the original detailed description, \textbf{Level~2} removes background and decorative details while preserving essential relations, and \textbf{Level~1} provides a concise description focusing on the primary subject and its most distinctive attributes.

\textbf{Verification and Refinement.}
In the final stage, all sketches and textual descriptions undergo expert verification to ensure semantic correctness, terminological consistency, and cross-modal alignment.
Domain experts in Thangka art and culture review the annotations against the standardized vocabulary and validate the correspondence among sketches, texts, and images.
All LLM-generated descriptions are manually verified and refined by domain experts to ensure semantic correctness and consistency.
This process improves annotation quality and helps ensure that CIRThan provides a robust benchmark for sketch+text composed image retrieval in knowledge-specific domains.

\subsection{Dataset Statistics and Analysis}

\subsubsection{Overview}
Table~\ref{tab:dataset_stats} summarizes the key statistics of CIRThan.
The dataset contains 2,287 Thangka images, each paired with one human-drawn sketch and three hierarchical textual descriptions, forming sketch--text--image triplets.
Following a standard train--test split, we allocate 80\% of the data for training and 20\% for test, resulting in 1,868 training triplets and 419 testing triplets.

The three textual levels are designed to represent progressively increasing semantic specificity.
As shown in Figure~\ref{fig:statistic}(\subref{text_length_box}), the average lengths of Level~1, Level~2, and Level~3 descriptions are 13.16, 23.10, and 34.13 words, respectively.
Importantly, this increase reflects controlled semantic enrichment rather than abrupt changes in linguistic structure, allowing systematic evaluation of CIR models under different degrees of semantic detail.
This hierarchical design supports systematic study of retrieval behavior under varying degrees of semantic granularity in composed queries.
Although CIRThan is relatively small in scale, it is designed for fine-grained, knowledge-intensive retrieval scenarios where annotation quality and semantic precision are more critical than data quantity.

\begin{table}[!t]
\centering
\caption{Summary statistics of CIRThan.}
\label{tab:dataset_stats}
\scalebox{1}{%
\begin{tabular}{lc}
\toprule[1pt]
\textbf{Property} & \textbf{Value} \\
\hline
\# Thangka Images & 2,287 \\
\# Sketches & 2,287 \\
\# Text Levels & 3 \\
\# Textual Descriptions & $2,287 \times 3$ \\
Train / Test split & 1,868 / 419 \\
Avg. image resolution & $1,700 \times 1,200$ \\
Avg. text length (Level 1 / 2 / 3) & 13.16 / 23.10 / 34.13 \\
\bottomrule[1pt]
\end{tabular}
}
\end{table}

\subsubsection{Scale Diversity of Thangka Imagery}
Figure~\ref{fig:statistic}(\subref{area_distribution}) compares the normalized distribution of image areas between CIRThan and the existing sketch+text CIR dataset CSTBIR~\cite{gatti2024composite}.
CIRThan exhibits a substantially broader distribution, spanning nearly four orders of magnitude in image area, with a median size of approximately $1.46 \times 10^{6}$ pixels.
By contrast, images in CSTBIR are concentrated around a much smaller median area of $1.87 \times 10^{5}$ pixels, resulting in a sharply peaked distribution.
This difference reflects the characteristics of Thangka arts, where high-resolution digitization is commonly adopted to preserve semantically meaningful details, such as intricate line work, dense decorative patterns, and layered compositional structures.
As a result, CIRThan includes significant variation in spatial scale, providing a realistic benchmark for studying CIR robustness under heterogeneous visual conditions commonly encountered in domain-specific and cultural heritage image collections.

\subsubsection{Semantic Diversity and Long-Tailed Subject Distribution}
Figure~\ref{fig:statistic}(\subref{tangka_top30_horizontal}) illustrates the distribution of visual subjects in CIRThan, which follows a pronounced long-tailed pattern.
While a small number of subjects appear frequently, a large portion of subject categories occur only a few times.
In particular, the aggregated ``Others'' category contains 267 distinct subject types and 789 images, accounting for approximately 34.5\% of the dataset.
This long-tailed distribution reflects the inherent semantic diversity of Thangka imagery, where imagery spans a wide range of rare and specialized subjects, and symbolic elements.
Such a distribution poses substantial challenges for retrieving Thangka images containing rare and specialized subjects, especially under strict semantic matching requirements, and makes CIRThan a realistic and domain-specific benchmark for studying CIR beyond dominant or frequently occurring semantic categories.

\section{Experiments}\label{sec:experiments}

\begin{table*}[!thbp]
\caption{Retrieval results on the CIRThan test set.
Supervised methods are trained on the CIRThan training split, whereas zero-shot methods are pre-trained on external image-text data or applied in a training-free manner.}
\begin{tabular}{l ccc ccc ccc ccc}
\toprule[1pt]
 \multicolumn{1}{l}{\multirow{2}{*}{\textbf{Method}}} & \multicolumn{3}{c}{\textbf{Level1}} & \multicolumn{3}{c}{\textbf{Level2}} & \multicolumn{3}{c}{\textbf{Level3}} & \multicolumn{3}{c}{\textbf{Average}}\\ 
 \cmidrule(lr){2-4}\cmidrule(lr){5-7}\cmidrule(lr){8-10}\cmidrule(lr){11-13}
& \textbf{R@1} & \textbf{R@3} & \textbf{R@5} &\textbf{R@1} & \textbf{R@3} & \textbf{R@5} & \textbf{R@1} & \textbf{R@3} & \textbf{R@5} & \textbf{R@1} & \textbf{R@3} & \textbf{R@5} \\ \hline
\multicolumn{13}{c}{\textbf{Supervised CIR Methods}} \\ \hline
Combiner~\cite{baldrati2022effective}& 25.30 & 47.97 & 57.28 & 27.68 & 50.84 & 60.14 & 30.07 & 54.42 & 66.83 & 27.68 & 51.08 & 61.42\\
Combiner-FT~\cite{baldrati2022conditioned} & 24.82 & 46.06 & 53.22 & 24.34 & 46.06 & 58.47 & 26.49 & 48.69 & 58.95 & 25.22 & 46.94 & 56.88\\
CLIP4Cir~\cite{baldrati2023composed} & 36.75 & 58.47 & 67.54 & 37.95 & 62.77 & 70.64 & 45.11 & 68.50 & 77.80  & 39.94 & 63.25 & 71.99\\
Blip4CIR~\cite{liu2024bi} & 24.58 & 40.10 & 49.10 & 27.92 & 53.22 & 62.77 & 29.59 & 55.13 & 63.01 & 27.36 & 49.48 & 58.29 \\
Bi-Blip4CIR~\cite{liu2024bi} & 26.25 & 45.82 & 54.42 & 28.40 & 51.31 & 62.05 & 28.40 & 52.51 & 61.81 & 27.68 & 49.88 & 59.43\\
CaLa~\cite{jiang2024cala} & 30.55 & 56.09 & 68.26 & 35.32 & 63.01 & 72.55 & 48.45 & 75.42 & 84.01 & 38.11 & 64.84 & 74.94\\
ENCODER~\cite{li2025encoder} & \textbf{49.88} & \textbf{73.75} & \textbf{81.38} & \textbf{51.55} & \textbf{73.51} & \textbf{83.53} & \textbf{54.65} & \textbf{79.24} & \textbf{87.59} &\textbf{52.03} &\textbf{75.50} &\textbf{84.17}\\ \hline
\multicolumn{13}{c}{\textbf{Zero-Shot CIR Methods}} \\ \hline
Sketch-only & - & - & - & - & - & - & - & - & - & 4.30 & 6.92 & 9.31 \\
Text-only & 6.21 & 12.17 & 16.71 & 5.97 & 13.60 & 19.57 & 7.16 & 16.95 & 20.53 & 6.44 & 14.24 & 18.94 \\
Pic2Word~\cite{saito2023pic2word} & 6.92 & 13.13 & 16.71 & 7.16 & 14.80 & 19.09 & 8.35 & 14.80 & 20.53 & 7.48 & 14.24 & 18.78 \\
LinCIR~\cite{gu2024language} & \textbf{7.88} & \textbf{17.90} & \textbf{21.00} & 7.40 & 15.99 & 21.24 & 8.59 & \textbf{17.90} & 23.15 & \textbf{7.96} & \textbf{17.26} & 21.80 \\
Context-I2W~\cite{tang2024context} & 7.63 & 15.75 & 20.29 & 7.64 & \textbf{17.42} & \textbf{24.11} & 8.35 & 16.47 & \textbf{24.82} & 7.87 & 16.55 & \textbf{23.07} \\ 
LDRE~\cite{yang2024ldre} & 6.68 & 15.27 & 20.29 & 6.92 & 16.95 & 20.53 & \textbf{8.83} & 16.47 & 20.29 & 7.48 & 16.23 & 20.37 \\
SEIZE~\cite{yang2024semantic} & 7.16 & 14.32 & 19.33 & \textbf{7.64} & 14.80 & 19.57 & 8.35 &15.27  & 20.05 & 7.72 & 14.80 & 19.65 \\
OSrCIR~\cite{tang2025reason} & 6.44 & 12.17 & 17.42 & 5.49 & 11.93 & 16.95 & 7.16 & 11.93 & 17.42 & 6.36 & 12.01 & 17.26 \\
CoTMR~\cite{sun2025cotmr} & 7.16 & 13.84 & 19.09 & 7.40 & 13.60 & 19.81 & 7.40 & 14.08 & 20.05 & 7.32 & 13.84 & 19.65 \\

\bottomrule[1pt]
\end{tabular}%
\label{baseline_results}
\end{table*}

\subsection{Baselines}
We evaluate representative CIR baselines under both supervised training and zero-shot transfer, to assess in-domain performance and cross-domain generalization on CIRThan.

\noindent \textbf{Supervised CIR.} 
We include widely adopted supervised CIR baselines trained with annotated triplets, including three variants of CLIP4Cir: Combiner~\cite{baldrati2022effective}, Combiner-FT~\cite{baldrati2022conditioned}, and CLIP4Cir~\cite{baldrati2023composed};
two variants of Bi-Blip4CIR: Blip4CIR and Bi-Blip4CIR~\cite{liu2024bi};
as well as CaLa~\cite{jiang2024cala} and ENCODER~\cite{li2025encoder}.

\noindent \textbf{Zero-shot CIR.}
We evaluate representative ZS-CIR methods that transfer pretrained vision-language models to CIRThan without using task-specific supervision.
These methods differ in whether they require lightweight adaptation or perform retrieval in a fully training-free manner.
\textbf{(i) Training-based ZS-CIR} methods require lightweight training on generic image-text data (without using CIRThan), including Pic2Word~\cite{saito2023pic2word}, LinCIR~\cite{gu2024language}, and Context-I2W~\cite{tang2024context}.
In contrast, \textbf{(ii) Training-free ZS-CIR} methods directly adopt LLMs/MLLMs to compose textual queries without any additional training, including LDRE~\cite{yang2024ldre}, SEIZE~\cite{yang2024semantic}, OSrCIR~\cite{tang2025reason}, and CoTMR~\cite{sun2025cotmr}.
In addition, we report sketch-only and text-only retrieval as reference settings by directly computing similarities in the CLIP embedding space.

% Although recent works~\cite{koley2024youll, gatti2024composite} have explored sketch+text CIR, direct comparisons are not included, as existing approaches lack publicly available and reproducible implementations, and this retrieval paradigm remains relatively underexplored, particularly in knowledge-specific domains.

\subsection{Experimental Settings}
\textbf{Dataset.}
All experiments are conducted on the proposed CIRThan dataset.
Each Thangka image corresponds to one human-drawn sketch and three hierarchical textual descriptions (Level~1-Level~3), which describe the same target image with increasing semantic specificity.
During training and evaluation, each sketch is paired with one text level at a time, forming three composed queries per target with different semantic granularity.
% During training and evaluation, we pair each sketch with one text level at a time, yielding three composed queries per image, each corresponding to a different level of semantic granularity.
This design allows us to systematically study how retrieval performance varies with the level of textual detail, while keeping the visual reference and target image fixed.
The dataset is split into 1,868 training images and 419 test images, and the corresponding sketch-text queries at all three levels are used consistently within each split.
All dataset splits and query constructions are publicly released to support reproducibility.
% All experiments are conducted on the proposed CIRThan dataset.
% Each Thangka image corresponds to a single human-drawn sketch and three hierarchical textual descriptions (Level~1-Level~3).
% During training and evaluation, a composed query is formed by pairing the sketch with one textual description at a time, with all levels corresponding to the same target image in the gallery.
% The dataset is split into 1,868 training triplets and 419 test triplets.

\noindent \textbf{Implementation details.}
All experiments are implemented in PyTorch~\cite{paszke2019pytorch} and conducted on a single NVIDIA RTX~4090 GPU.
Baseline models follow the default settings provided in their original implementations, without dataset-specific hyperparameter tuning.
Supervised CIR models are trained on the CIRThan training set and evaluated on the test set.
For zero-shot methods, training-based ZS-CIR approaches follow their original training protocols and are trained on large-scale generic image-text datasets (e.g., CC3M~\cite{sharma2018conceptual}), without using any CIRThan training data, and are then evaluated on the CIRThan test set.
Training-free ZS-CIR baselines rely on pretrained MLLMs without task-specific training.
We adopt Qwen2-VL-7B~\cite{Qwen2-VL}, as the default MLLM for all training-free baselines, unless otherwise specified.

% For zero-shot methods, we strictly follow the original training protocols of each approach.
% Specifically, training-based ZS-CIR baselines~\cite{saito2023pic2word, gu2024language, tang2024context} are trained on large-scale generic image-text datasets (e.g., CC3M~\cite{sharma2018conceptual}) as described in their original papers, and are directly evaluated on the CIRThan test set without using any CIRThan training data.
% Training-free ZS-CIR baselines rely on pretrained LLMs or MLLMs without task-specific training.
% In our experiments, we adopt Qwen2-VL-7B as the default MLLM for all training-free baselines, unless otherwise specified.

\noindent\textbf{Evaluation Metrics.}
Following common practice in composed image retrieval, we evaluate all methods using Recall@K (R@K).
For each composed query, all Thangka images in the gallery are ranked by similarity, and retrieval is considered successful if the ground-truth target Thangka appears within the top-K results.
We report R@K for $K \in \{1, 3, 5\}$, as each query in CIRThan has a single ground-truth target.

\subsection{Results}
We report the retrieval performance of representative supervised and zero-shot CIR baselines on CIRThan in Table~\ref{baseline_results}.
Overall, supervised methods substantially outperform zero-shot counterparts, and retrieval accuracy consistently improves with richer textual descriptions.
Our analysis focuses on how existing CIR paradigms behave in a culturally grounded retrieval scenario that requires domain-specific knowledge and fine-grained matching between sketch+text composed queries and target Thangka images.

%表1展示了监督学习方法在唐卡测试集上的表现。这些方法都在唐卡训练集上训练，然后在测试集上验证结果。在所比较的基线中，CLIP4CIR 的平均 R@1 达到 39.94%，显著高于 Combiner（27.68%）、Combiner-FT（25.22%）、Blip4CIR（27.36%）以及 Bi-Blip4CIR（27.68%）。这一结果表明，在唐卡这种与通用真实世界数据集（如 CIRR）存在显著分布差异的场景中，仅对文本侧或融合模块进行轻量适配往往不足；对视觉表示进行更充分的域内适配（在我们的训练配置下，CLIP4CIR 具备更强的视觉侧更新能力）能够带来更明显的性能收益。其原因在于唐卡图像具有更强的文化风格特征与更高的视觉复杂度（纹样密集、装饰元素多、背景干扰强），模型需要通过域内训练才能学到更符合唐卡判别需求的视觉线索。
%更进一步，ENCODER 在平均 R@1 上达到 52.03%，相比 CLIP4CIR 提升 12.09 个百分点，取得表1中最优结果。我们认为该显著提升不仅来自域内训练本身，更来自其对细粒度修改线索的更强建模能力。相比于以全局对齐/全局融合为主的基线，ENCODER 更强调从图文交互中挖掘实体和实体之间的修改关系。由于唐卡数据集中大量样本在全局构图上高度相似，而关键差异往往体现在局部实体细节及其关系上（如手持物、手势/动作、局部颜色与装饰层级等），因此具备更强局部与关系表征能力的方法在唐卡数据集上更具优势。
\subsubsection{Supervised CIR on CIRThan.} 
Supervised CIR methods trained on CIRThan achieve markedly stronger performance than zero-shot baselines.
As shown in Table~\ref{baseline_results}, ENCODER attains the highest average R@1 of 52.03\%, outperforming CLIP4Cir (39.94\%) and other CIR methods by a large margin.
Across all supervised models, retrieval accuracy consistently increases from Level~1 to Level~3.
For instance, ENCODER improves from 49.88\% at Level~1 to 54.65\% at Level~3 in R@1.
This trend indicates that richer textual descriptions provide increasingly informative semantic constraints for distinguishing visually similar Thangka images.
Overall, although supervised models benefit from in-domain training, their performance remains far from satisfactory, highlighting the intrinsic difficulty of fine-grained semantic matching in CIRThan, where many images share similar global composition and discriminative cues are embedded in subtle attributes, symbolic elements, and structured semantic relations that require fine-grained understanding.

% This gain suggests the importance of capturing fine-grained entities and their relations in Thangka retrieval: many images are globally similar, while the discriminative differences often lie in local details and interactions (e.g., gestures/actions, hand-held objects, and local color/ornament patterns). Compared to baselines that primarily perform global matching, ENCODER emphasizes mining entities and modification relations through image–text interaction, which is better aligned with the characteristics of the Thangka dataset.

%table2：表2展示了常见的 zero-shot CIR 方法在唐卡测试集上的表现。观察表中数据发现，单模态检索的方法（仅草图检索和仅文本检索）的结果低于zero-shot上组合检索的结果，sketch-only和text-only的averageR@1分别为4.3\%,6.44\%。这主要是由于草图提供位置信息，文本提供细节信息，二者对于唐卡图片来说是互补的，因此单模态的检索会低于多模态的检索效果。此外，我们发现文本反转的CIR方法的结果基本和免训练的CIR方法持平，例如文本反转方法里表现较好的LinCIR的average R@1是7.96%，而免训练的方法中表现较好的LDRE和SEIZE的average R@1分别为7.48%和7.72%（如果后面再做GPT实验的话需要改动）

\subsubsection{Zero-shot CIR on CIRThan.}
Zero-shot CIR methods exhibit substantially lower performance than supervised models.
The strongest zero-shot baseline achieves an average R@1 below 8\%, compared to over 50\% for supervised methods, revealing a pronounced domain gap between generic image–text training data and Thangka imagery.
Single-modality baselines perform particularly poorly, with sketch-only and text-only retrieval achieving average R@1 scores of 4.30\% and 6.44\%, respectively.
This observation confirms that neither modality alone is sufficient to capture the complex semantic information of Thangka images, which often require both structural visual cues and explicit textual specification.

Although multi-modal zero-shot methods consistently outperform single-modality settings, their overall accuracy remains limited.
For instance, LinCIR achieves the highest zero-shot average R@1 of 7.96\%, while training-free approaches such as SEIZE and LDRE yield comparable results.
These findings indicate that existing ZS-CIR methods struggle to achieve fine-grained alignment between composed sketch+text queries and Thangka without in-domain supervision, underscoring CIRThan as a challenging benchmark for cross-domain generalization.

\subsubsection{Impact of textual granularity.}
Across both supervised and zero-shot settings, retrieval performance generally improves as textual descriptions become more detailed.
More detailed descriptions provide richer semantic specification, which helps disambiguate visually similar Thangka images, particularly when differences lie in attributes, symbolic elements, or subtle relational cues.
Importantly, this trend is consistent across different CIR paradigms, suggesting that hierarchical textual specification is a fundamental factor for composed image retrieval in knowledge-specific domains.
These findings validate the design choice of CIRThan to include multi-level textual descriptions and highlight the importance of modeling semantic granularity when studying fine-grained CIR.

\begin{table}[!t]
\caption{The performance of different MLLMs of CoTMR on CIRThan test set.}
\centering
\scalebox{1}{%
\begin{tabular}{lcccc}
\toprule[1pt]
\multirow{2}{*}{\textbf{Method}} & \multirow{2}{*}{\textbf{MLLM}} & \multicolumn{3}{c}{\textbf{Level3}} \\ \cline{3-5} 
 & & \textbf{R@1} & \textbf{R@3} & \textbf{R@5} \\ \hline
\multirow{4}{*}{CoTMR} &
Qwen2-VL-7B & 7.40 & 14.08 & 20.05 \\
& Qwen2.5-VL-72B & 7.40 & 15.99 & 21.00 \\
& GPT-4o mini & 8.35 & 14.56 & 19.57 \\
& GPT-4.1 & 7.40 & 15.27 & 21.96 \\
\bottomrule[1pt]
\end{tabular}%
}

\label{comparison_LLMs}
\end{table} 

\subsubsection{Analysis of Different MLLMs.} 
Table~\ref{comparison_LLMs} compares the performance of different multimodal large language models~(MLLMs) when used within the CoTMR framework.
We select four representative MLLMs: Qwen2-VL-7B~\cite{Qwen2-VL}, Qwen2.5-VL-72B~\cite{bai2025qwen2}, GPT-4o mini~\cite{openai_gpt4omini_2024}, GPT-4.1~\cite{openai_gpt41_2025}. 
Despite substantial differences in model scale and pretraining data, retrieval performance varies only marginally across MLLMs.
These results suggest that current MLLMs, even at larger scales, remain limited in their ability to jointly reason over sketch-based visual abstractions and domain-specific textual descriptions for CIR.
The challenge appears to stem not from language modeling capacity alone, but from the difficulty of grounding composed sketch+text queries in culturally specific visual semantics.
Overall, these observations further emphasize that CIRThan poses challenges that go beyond generic multimodal understanding, requiring advances in cross-modal fine-grained alignment, sketch understanding, and domain-aware reasoning.

% Table~\ref{tab:4} reports the performance of different multimodal large language models (MLLMs) when used within the CoTMR framework.
% We select four representative MLLMs: Qwen2-VL-7B~\cite{Qwen2-VL}, Qwen2.5-VL-72B~\cite{bai2025qwen2}, GPT-4o mini~\cite{openai_gpt4omini_2024}, GPT-4.1~\cite{openai_gpt41_2025}. 
% As shown in the table, changing the underlying MLLM yields only marginal differences in retrieval accuracy, even across models with substantially different parameter scales.
% These results suggest that current MLLMs still face challenges in effectively understanding and integrating sketch and text queries for retrieval in complex and densely structured visual domains such as thangka imagery.

\section{Potential Research Directions}
CIRThan is designed to advance CIR in culturally grounded and knowledge-specific visual domains by integrating free-form sketches, hierarchical texts, and Thangka images under fine-grained matching requirements.
It enables several research directions that are insufficiently supported by existing CIR benchmarks.

\textbf{(1) User-centric sketch+text CIR.}
Sketch+text CIR remains an early-stage research direction with limited benchmarks.
In CIRThan, sketches provide a flexible and intuitive way for users to express fine-grained visual intent through selective abstraction and visual cues, while hierarchical textual descriptions explicitly model variations in the complexity and specificity of user retrieval intent.
This design enables systematic study of user-centric multimodal query formulation and composition, as well as how retrieval models integrate complementary visual and textual semantics under realistic intent expression.

\textbf{(2) Knowledge-specific retrieval with dense and structured semantics.}
Thangka imagery contains multiple semantically meaningful entities, symbolic objects, and culturally grounded configurations organized in highly structured compositions.
CIRThan therefore supports research on composed retrieval beyond coarse attribute modification, requiring models to align sketch+text queries with instance-level attributes, symbolic details, and their structural relations.
This setting enables investigation of fine-grained cross-modal matching in knowledge-specific visual domains.

\textbf{(3) Limitations of MLLMs in knowledge-specific domains.}
Experimental results on CIRThan reveal a substantial performance gap between supervised methods and zero-shot approaches based on MLLMs.
Despite their strong general vision–language understanding capabilities, current MLLMs (e.g., Qwen-VL, GPT-based models) struggle to achieve reliable semantic alignment between sketch-based abstractions, domain-specific textual semantics, and fine-grained Thangka images.
This suggests that existing MLLMs have limited capacity to interpret non-natural visual inputs such as sketches and to integrate culturally grounded domain knowledge under fine-grained matching requirements.
CIRThan therefore motivates research on improving MLLM-based retrieval in knowledge-specific domains, including domain adaptation, knowledge-aware representation learning, and retrieval-augmented reasoning.

\section{Conclusion}\label{sec:conclusion}

In this work, we introduced \textbf{CIRThan}, a sketch+text composed image retrieval dataset for Thangka imagery, a knowledge-specific visual domain where successful retrieval requires fine-grained semantic alignment between composed sketch+text queries and visually complex images.
CIRThan consists of 2,287 high-quality Thangka images, each paired with a human-drawn sketch and hierarchical textual descriptions at three semantic levels, enabling expressive and structured multimodal queries.
We provide standardized data splits, comprehensive dataset analysis, and benchmark evaluations of representative supervised and zero-shot CIR methods.
Experimental results show that existing CIR approaches, largely developed on general-domain imagery, face substantial challenges when applied to sketch+text queries and knowledge-specific domains, particularly in the absence of in-domain supervision.
We believe CIRThan offers a valuable benchmark for advancing research on sketch+text composed image retrieval, hierarchical semantic query modeling, and multimodal retrieval in cultural heritage and other knowledge-specific visual domains.

% In this work, we introduced CIRThan, a sketch+text composed image retrieval dataset for Thangka imagery.
% CIRThan consists of 2,287 high-quality Thangka images, each paired with a human-drawn sketch and hierarchical textual descriptions at three semantic levels, enabling expressive multimodal queries in a knowledge-specific visual domain.
% In addition to dataset construction, we provided standardized data splits, comprehensive statistical analysis, and benchmark evaluations of representative supervised and zero-shot CIR methods.
% Our experimental results indicate that existing CIR approaches, which are primarily developed and evaluated on general-domain imagery, face notable challenges when applied to sketch+text queries and knowledge-dependent visual content.
% In particular, effectively aligning sketches and hierarchical textual descriptions with culturally grounded Thangka images remains difficult without in-domain supervision.
% We believe CIRThan offers a valuable resource for advancing research on sketch+text composed image retrieval, hierarchical semantic query understanding, and multimodal retrieval in cultural heritage and knowledge-specific visual domains.

%%
%% The acknowledgments section is defined using the "acks" environment
%% (and NOT an unnumbered section). This ensures the proper
%% identification of the section in the article metadata, and the
%% consistent spelling of the heading.
\begin{acks}
This research is partially supported by National Natural Science Foundation of China (Grant No. 62271360) and National Key Research and Development Program of China (2024YFF0907002)
\end{acks}

\clearpage
%%
%% The next two lines define the bibliography style to be used, and
%% the bibliography file.
\bibliographystyle{ACM-Reference-Format}
\balance
\bibliography{sample-base}

\end{document}